# Assessing a Safety Case: Bottom-up Guidance for Claims and Evidence Evaluation

Schnelle S.[a], Favarò F.M.[a], Fraade-Blanar L.[a], Wichner, D.[a], Broce H.[a], Miranda J.[b]
[a] Waymo LLC, [b] Pyramid Consulting

**Abstract:** As Automated Driving Systems (ADS) technology advances, ensuring safety and public trust requires robust assurance frameworks, with safety cases emerging as a critical tool toward such a goal. This paper explores an approach to assess how a safety case is supported by its claims and evidence, toward establishing credibility for the overall case. Starting from a description of the building blocks of a safety case (claims, evidence, and optional format-dependent entries), this paper delves into the assessment of support of each claim through the provided evidence. Two domains of assessment are outlined for each claim: procedural support (formalizing process specification) and implementation support (demonstrating process application). Additionally, an assessment of evidence status is also undertaken, independently from the claims support. Scoring strategies and evaluation guidelines are provided, including detailed scoring tables for claim support and evidence status assessment. The paper further discusses governance, continual improvement, and timing considerations for safety case assessments. Reporting of results and findings is contextualized within its primary use for internal decision-making on continual improvement efforts. The presented approach builds on state of the art auditing practices, but specifically tackles the question of judging the credibility of a safety case. While not conclusive on its own, it provides a starting point toward a comprehensive "Case Credibility Assessment" (CCA), starting from the evaluation of the support for each claim (individually and in aggregate), as well as every piece of evidence provided. By delving into the technical intricacies of ADS safety cases, this work contributes to the ongoing discourse on safety assurance and aims to facilitate the responsible integration of ADS technology into society.

---

## 1. Introduction

Level 4 Automated Driving Systems (ADS) (SAE, 2021), which require no intervention by a human driver, present a transformative opportunity to enhance transportation safety, efficiency, and accessibility. However, the complexity of these systems introduces unique safety challenges that necessitate robust and adaptable assurance frameworks. As ADS technology progresses towards widespread deployment, establishing public trust and regulatory acceptance hinges on the ability to demonstrate that these systems are acceptably safe in a convincing manner.

In recent years, safety cases have emerged as a viable approach to address those challenges, supported by industry and regulators alike (UL, 2023; ISO, 2025; EU, 2022; UNECE, 2025). The use of safety cases traces back to a number of diverse, safety critical industries, including railway (HSE, 1994), maritime (UK MoD, 1996), chemical (Cullen, 1990), and aerospace (Gheman et al., 2003), especially in connection with software-intensive systems (Barker et al., 1997). Informed by this rich history, safety assurance use cases have been formalized into the well known definition for a safety case:

> ***"A structured argument supported by a body of evidence that provides a compelling, comprehensible, and valid case that a system is safe for a given application in a given environment"*** (UK MoD, 2017; UL, 2023).

In 2023, Waymo presented its approach to the creation of a safety case (Favarò et al., 2023), detailing the importance of tackling the question of determining "absence of unreasonable risk" (which is the standardized definition of safety (ISO, 2018)) in a manner that sufficiently demonstrates competence to operate on public roads. That presentation was grounded in standardized definitions of acceptance criteria and validation targets (ISO, 2022), detailing a theoretical framing to encompass a layered hazard-based risk assessment within the dynamic and iterative development of a software release process. Other ADS companies also shared information about their approach to safety, including details on the structure and topics of their safety case (Aurora, 2025), and/or commitment for validation of the safety determination (see Gatik's [Safety Case Assessment Framework](#)).

It can be argued that safety cases have the potential to become a central safety assurance artifact for ADS developers to provide evidence of a mature approach to ensure safety, with the intent of offering a more transparent and comprehensive framing for the evaluation and acceptance of these novel systems. Yet, the presentation of a structured argument is not, by itself, a sufficient indicator of the engineering rigor supporting the safety determination. In fact, following the careful curation of evidence in support of each claim in the safety case, a precise *assessment process* that evaluates the support of each claim has to be executed. This process continues to evolve as more experience is gained in evaluating safety cases.

Note that the process described here is focused on an ADS developer's independent self-assessment of its safety case, where independence is achieved by having a separate team of safety experts performing the assessments who are not directly involved in the development or evaluation of the ADS. This independent internal assessment seeks to examine every claim and its supporting evidence to evaluate (e.g., via scoring) each of those elements to determine their soundness. Deficiencies revealed by this process would lead to improvements in the safety case. By contrast, assessment of an ADS safety case by or for a regulatory agency (e.g., in connection with type approval or a review of self-certification of an ADS) may employ a somewhat similar process but would most likely be focused more broadly on determining the safety case's overall credibility by examining the objective completeness of the claims with regard to the sufficiency of their coverage of the ADS's operational design domain and the pertinence of the evidence provided in support of the claims along with clear indications that the evidence was derived through acceptable means of testing or analysis. Regulators may require ADS manufacturers to conduct an independent assessment of the type addressed herein and provide a report of the independent assessment and resultant remedial actions as part of a safety case submission.

**This paper details how the claims and evidence assessments were undertaken on behalf of an ADS developer in an actual Level 4 ADS safety case for operation without a human driver.** A number of standards (ISO, 2019; ISO, 2015; VDA, 2020; UL, 2023) can support how

such an assessment takes place, starting from the evaluation of each evidence piece leveraged within the safety case and then evaluating the support of each individual claim. The process in this paper builds on these standards to help answer the questions of how to evaluate the credibility of a safety case, part of what was termed Case Credibility Assessment (CCA)(Favarò et al., 2023). The CCA rests on the top-down pillar of *credibility of the argument* and the bottom-up pillar of *credibility of the evidence*, further reinforced by an *implementation credibility* check. Previous literature, including the aforementioned auditing standards and industry frameworks, lacked a detailed process for how the bottom-up portion of this assessment is conducted and operationalized towards the safety assurance of an ADS. To address this need, this paper presents a process for the assessment of an individual claim's support through the provided evidence, as well as the assessment of each piece of evidence. These assessments are a required process that needs to be applied during the evaluation of a safety case by or on behalf of an ADS developer. The top-down portion of the assessment regarding the sufficiency and completeness of a safety case's arguments is out of scope for this paper.

In this paper, we begin with a presentation of the building blocks for any safety case: claims that build up an overarching argument for safety; supporting evidence; and other (optional, and format-dependent) entries. Following precise definitions, we outline the steps for the creation of a safety case, within a discussion on governance, continual improvement, and timing considerations. We further explore the assessment phase for the support of the safety case, where two distinct domains of assessment are detailed: 1) the assessment of each claim's support through the provided evidence, and 2) the assessment of each piece of evidence overall status (e.g., recency) and documentation control (e.g., oversight on modifications). Across the assessment approach, we distinguish between procedural support and implementation support (discussed in detail in Section 4), creating a holistic framing for the safety case. The execution of such an assessment leads to the systematic evaluation and aggregation of results for each branch of the safety case. Scoring strategies and evaluation guidelines are also presented in this paper, along with describing the technical intricacies of safety cases for ADS, exploring structure, assessment methodologies, and role within a broader safety management system (SMS). Through a detailed examination of key concepts and practical examples, this paper aims to contribute to the operationalization and technical rigor of safety assurance for ADS to ensure their responsible and beneficial integration into society.

## 2. Anatomy of a Safety Case

The purpose of a safety case is to provide a credible argument that an ADS developer's safety approach is mature and that their system is acceptably safe. Industry standards that discuss the notion of a safety case are not prescriptive in what specific type of argumentation approach should be included, nor do they denote the breadth or depth of the arguments used. Rather there is general consensus around the fact that the first step in development of a safety case is the definition of the safety case boundaries, or, in other words, the scope of the safety case.

Looking back at the definition provided in the Introduction, we can parse the following elements:

- *"[...] a system [...]"*: a safety case focuses on one particular system. While a developer or company may be concurrently running multiple projects at the same time, the safety case is specific to a single system. This can entail, for example, specific architectural choices that differentiate tiers of similar products. The system definition also speaks to whether the subject of the safety case is a complete system, a sub-system, or even a single component. This distinction can have important consequences on the particular formatting and language chosen to detail the safety case argument structure, where safety case examples abound in the literature for subsystem and component level applications (see for example (Salay et al., 2021) and references therein), but are scarcer for system (or system of systems) level applications. In general, literature points to a correlation between the formality of the languages used to build the safety case and the level of abstraction of the item under analysis: for example, more formal (e.g., mathematical and/or logic-based symbolism) languages can be often found in component-level applications (e.g., perception function only through one particular sensing solution), while natural english language is favored for system-level applications.
- *"[...] for a given application [...]"*: this entry describes the use cases or usage specification for the system. The definition of a use-case is necessary to identify and narrow down the potential hazards and risks that the safety case should speak to. For example, the specification of a Level 4 ADS (SAE, 2021) for ride-hailing would point to the need for in-depth analysis of the role of riders in the vehicle (e.g., safety analysis of passengers pick-ups and drop-offs); conversely, a use-case featuring goods delivery in a commercial vehicle would exclude those considerations, while focusing more, for example, on stability and control connected with cargo loading and securement.
- *"[...] in a given environment [...]"*: this entry is grounded in the requirements and conditions that enable operations for the system. For ADS applications, this includes the specification of the Operational Design Domain (ODD), which specifies external conditions like environmental, geographical, and time-of-day conditions (SAE, 2021) under which an ADS is designed to function. Similar to the usage specification, the ODD also drives which specific attributes need to be considered to ensure proficiency in performing the Dynamic Driving Task (DDT, (SAE, 2021)), including the Object and Event Detection and Response (OEDR, (SAE, 2021)) capabilities for the system.

The definition of the item under analysis for a safety case (inclusive of the system overview, the use case definition, and the specification of the operational design conditions) determines the scope and breadth of the safety case, but it historically has not been considered part of the safety case itself. More often than not, it will be included as front matter and/or supplemental information to the structured safety case itself.[1]

---

[1] This front matter could be found in what is sometimes referred to as a "safety case report". Additional content, such as an explanation of the developer's approach to the safety case creation (see for example the content in (Favarò et al., 2023)), can help reviewers better understand the structure and breadth of the content included in the safety case. The creation of a safety case report, which outlines and breaks-down the safety case content, facilitates higher level understanding of the system and the argumentation approach used in the safety case for external reviewers. This report may also overview the engineering methodologies used to evaluate the system performance and the overall safety concept —a description of the measures integrated into a system's design to ensure its operation is free from unacceptable safety risks to users and others within the intended operational domain. Specific measures implemented to achieve the system's safety requirements, including redundant systems, fail-safe mechanisms, emergency procedures and fallback strategies can also be included and we expect that the detailed content and outline of these types of reports will be shaped by future regulatory requirements.

Another area where general consensus exists is connected to the high-level structure for crafting a logical argument. Literature (Dezfuli, 2015; ISO, 2019; MISRA, 2019) recognizes two elements that are at the basis for constructing a safety case: *claims* and (supporting) *evidence*. As noted below, additional elements and/or refinements of the notion of "claim" are format-specific.

## *2.1 Claims*

At the heart of a safety case are the claims that build up the safety cases structure, or "**A structured argument [...]**" from the safety case definition. We define a claim as an *assertion that supports the overarching safety determination and whose veracity can be confirmed and/or verified.* In other words, claims are falsifiable statements (UL, 2023).

Claims represent key objectives and properties that the safety case aims to demonstrate. They give the safety case shape and purpose. Non-exhaustive characteristics of a strong claim include it being (Dezfuli, 2015; ISO, 2019; SCSC, 2021):

- Specific: clearly stating what aspects of the system are being covered by the claim and under what conditions, avoiding overstatements such as all, any, every;
- Verifiable: formulating claims in a manner that allows, whenever possible, for objective testing, measurement, verification, and/or confirmation of assumptions;
- Representative: stating claims in a realistic way that is representative of the system's actual design and operational context;
- Contextualized: formulating claims according to clearly defined boundaries of the system, its intended operational environment, and any relevant assumptions.

Within a safety case, a claim is seldom drafted in isolation. In fact, the safety case itself is built around the decomposition of a top-level goal (or primary claim) into hierarchical sub-claims. This is intuitively represented in various graphical notations highlighting said decomposition, as presented in Figure 1 for a higher level claim (referred to as a "parent claim") further decomposed into two supporting sub-claims (referred to as "child claims").

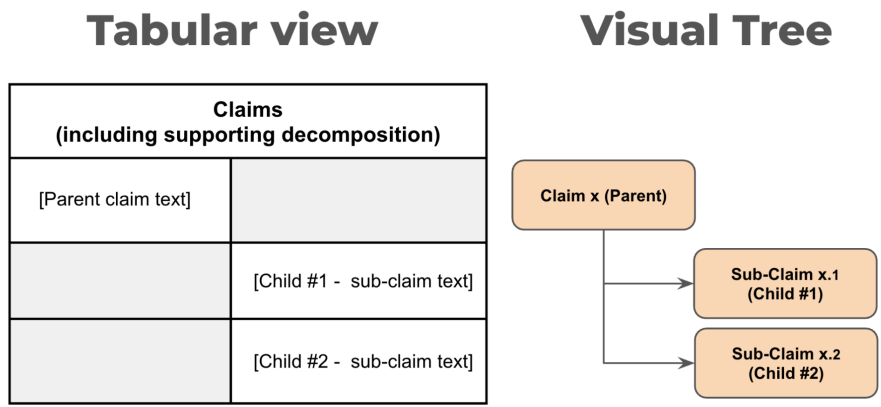

Figure 1. Visual decomposition of a parent claim into two child sub-claims in both tabular format and through a visual tree.

The top-level claim of a safety case (often also referred to as top-level goal) is usually centered on the declaration of the *system under analysis being safe*. This corresponds to the " ***[...] is safe [...]***" portion of the definition, where a number of options (both technical and philosophical) exist on the *principles and/or criteria that would underpin the definition of safety* (Favarò et al., 2023; ISO, 2025; Favarò, 2021; ISO, 2022; Koopman, 2022). The selection of such criteria in turn grounds the decomposition of the top-level claim into sub-claims of interest (at times referred to as "branches" of a safety case due to the decomposition shown in Figure 1). This selection and the resulting decomposition lead to the uniqueness of each individual developer's safety argumentation. Within Waymo's approach, the *top-level goal of absence of unreasonable risk* is decomposed into an array of supporting acceptance criteria, one for each of the internal engineering methodologies that form the basis for the evaluation of a candidate software release toward deployment approval (Favarò et al., 2025). For each criterion, the further decomposition follows a binary structure presented in Favarò et al. (2023), where a first sub-claim (1. Reasonableness of acceptance criterion in Figure 2) pressure-tests the reasonableness of the criterion chosen and a second sub-claim (2. The methodology provides credible evidence for criterion evaluation in Figure 2) pressure-tests the ability of a given engineering methodology to appropriately evaluate such criterion. This structure is presented in Figure 2, together with additional decomposition details that grounded the creation of Waymo's safety case.

The logical decomposition of claims into supporting sub-claims requires a balancing act between the sought level of detail of the safety case and the robustness/survivability to inevitable changes within the system and/or its supporting processes. In fact, decomposition into too granular claims may entail the need for frequent updates and hamper long-term survivability of the safety case. Care should be exercised into finding a level of abstraction that supports logical validity and understanding of the safety case, while also retaining the minimum salient aspects. This is a well known trade-off, as called out in MISRA (2019).

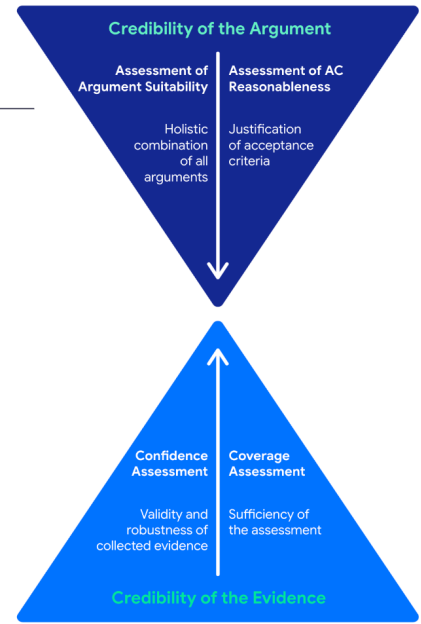

Figure 2. Overview of the decomposition structure employed for each acceptance criterion within Waymo's safety case. Adapted from (Favaró et al., 2023).

## *2.2 Evidence*

Evidence provides artifacts toward satisfaction of the safety case claims, as shown in the text "*[...] supported by a body of evidence [...]*" from the definition. Evidence can be defined as *any artifact establishing useful facts to substantiate a claim*. Evidence artifacts can take many shapes and forms: a process-flow document; results from a testing activity; a dashboard monitoring live data from the system; etc.. In general, we distinguish between two macro categories of evidence (discussed in greater detail in section 4.1):

i. Evidence providing *"procedural support"* to a claim, intended as an artifact that addresses a process (or product design) specification;
ii. Evidence providing *"implementation support"* to a claim, intended as an artifact that addresses the application of a process, providing results and/or example use-cases.

In other words, do we say what we do, and do we do what we say (with "we" as the primary drafters of the safety case). For both categories, characteristics of strong evidence towards substantiating a claim include:

- Relevancy: directly addressing the specific safety aspects outlined in the claims, avoiding extraneous or tangential information.
- Validity: originating from credible sources and through rigorous methods, such as controlled experiments, validated simulations, or statistical analysis of operational data.
- Traceability: being clearly documented and linked to the claim, allowing for verification and independent assessment.

Evidence is appended to the related claims, as intuitively visualized below in Figure 3, following the same structure presented before in Figure 1.

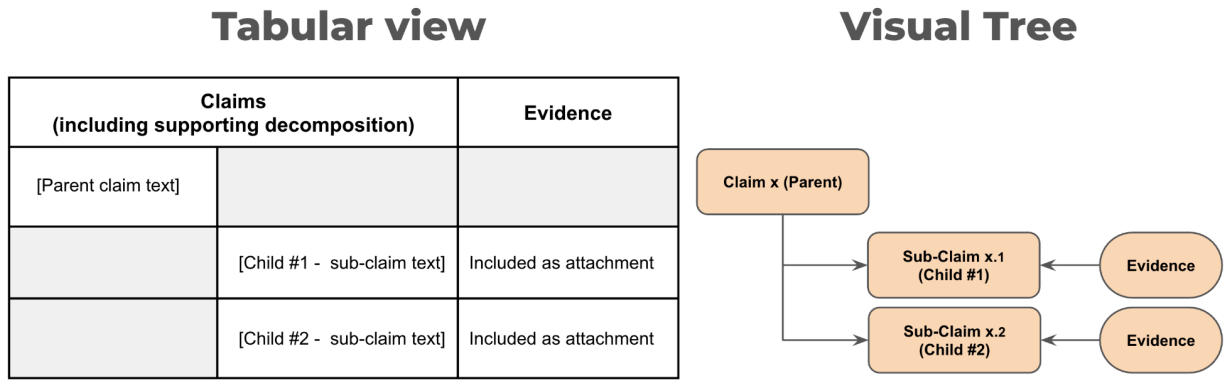

Figure 3. Visual decomposition of a parent claim into two child sub-claims in both tabular format and through a visual tree. Evidence is appended to each of the child claims in support of the parent claim.

Similar to the trade-offs happening for claims, collecting evidence in support of the safety case requires a careful balance between over-inclusion and sufficient support of the claim. Out of an abundance of caution, one could be tempted to provide any artifact more or less closely connected to the claim. While over-inclusion of evidence does not harm the logical validity of a claim, it can in fact hamper the long-term maintainability and survivability of the safety case as well as the ability to discern critical evidence from the crowd. Additionally, "quantity" of evidence does not necessarily speak to satisfactory substantiation of the claim, so that multiple pieces of evidence could still fail to provide adequate support of core elements of a claim. This will be a central aspect in the claim assessment guidelines we discuss later in this paper.

### *2.3 Argument: The (missing) definition*

Looking back at the safety case definition, the only remaining portion to be covered is its qualification to provide "***[...] a compelling, comprehensible, and valid case [...]***". This statement lives in a different stage or phase of the safety case: it speaks to the evaluation of the case merits rather than its creation. It is, in other words, connected to the *assessment* of a completed safety case, rather than the drafting of claims and the collection of evidence. In relation to the Case Credibility Assessment (CCA) mentioned in the Introduction and presented in Favaró et al., (2023), this qualification applies to both the top-down and the bottom-up aspects of the CCA: that is, an analysis of "compelling, comprehensible, and valid" characteristics of a safety case applies both to individual claims in the safety case (i.e., is each claim sufficiently supported by the provided evidence?) and to the resulting totality of the safety case body (i.e., are the provided claims and evidence appropriate and sufficient to support the top-level goal?). Within this paper, we tackle the bottom-up portion of the assessment, and leave the top-down approach as out-of-scope (though further considerations are provided in the Limitations section below). Additional details regarding aspects of the top-down approach can be found in Favaró et al. (2021;2025). Both are presented in Figure 4 below, outlining how the decomposition of individual safety case branches (as presented in Figure 2 above) connects with the different levels of abstraction of the CCA within Waymo's approach.

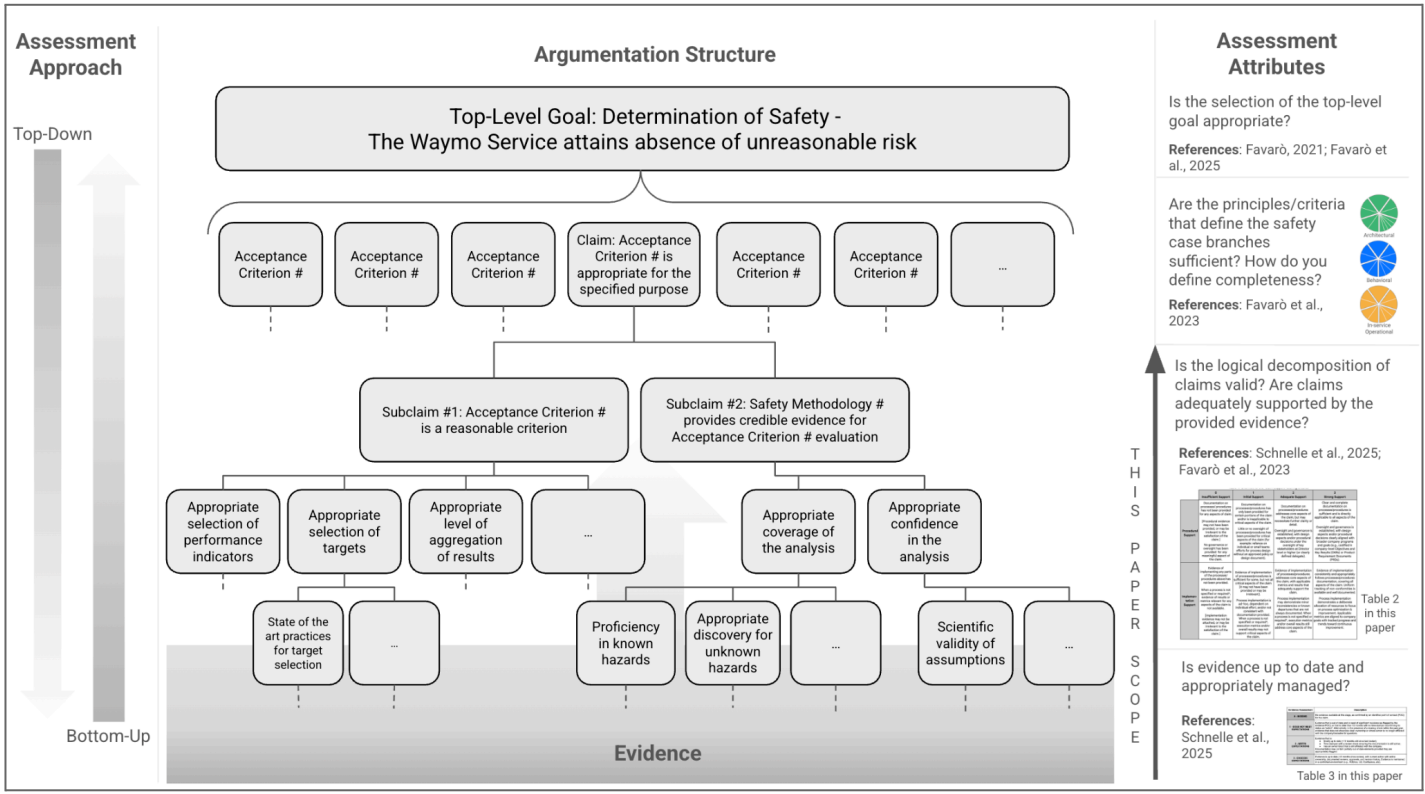

Figure 4. Connection between the assessment abstraction level, the argumentation structure, and the attributes in scope for each assessment level.

For both the bottom-up portion of the assessment (i.e., checking the support of each individual claim and evaluating the connections between hierarchical claims' decompositions) and the top-down portion (i.e., evaluating how comprehensive and robust the entirety of the case is), a third element of a safety case can be used: the argument. As noted in the beginning of this section, scientific literature points to broad consensus around the central role of *claims* and *evidence*, with clear and mature definitions for such elements. The same cannot be necessarily said for the notion of "argument", where many standards use this term based on its common English-vocabulary meaning, without providing for it a further technical definition. In fact, the argument surfaces, perhaps implicitly, from the decomposition of the top-level claim (the branches we referenced before), so much so that oftentimes people intuitively discuss the notion of argument as a synonym to the whole safety case itself ("***A structured argument […]***").

Within the literature, individual sub-claims that are supported by evidence are sometimes referred to as "atomic arguments". The combination of more atomic arguments in turn builds a "compound argument" toward satisfaction of a parent claim (Holloway, 2021; UL, 2023). Embedded in the differentiation between atomic and compound arguments is the (design) decision of whether to opt to further decompose a claim, or stop and provide supporting evidence. For example, within Figure 3 a parent claim is supported by a decomposition into two child claims, each supported by evidence. We could conceive that the same evidence could have been appended to the parent claim and still lead to the same level of support for it. This can create a trade-off when considering sufficient decomposition between the specificity of atomic claims and the practical desire to contain the total number of claims: on one hand, narrowly-scoped atomic claims point to more specific evidence and simplify the assessment of their support; on the other hand, a too-broad of a scope for a claim may increase the subjectivity in the assessment of the provided evidence sufficiency and applicability. The decision of *what level of claim to provide evidence* and *what evidence to provide* speaks to an argumentation strategy, and, specifically, bottom-up characteristics for it. Guidelines for informing this decision stem from ISO (2019) and MISRA (2019) and include:

- Clearly linking relevant evidence or sub-claims to claims: explicitly demonstrate how the presented evidence or sub-claims supports the specific safety claim being made, minimizing logical gaps and maximising traceability.
- Addressing potential (dialectic) weaknesses: acknowledge any uncertainties, limitations or potential counter-arguments to the evidence or claim itself. The argument should then provide justifications, counter-arguments, additional evidence to mitigate concerns, or note potential weaknesses.
- Being logically sound: employ valid reasoning, explicitly state assumptions, and avoid fallacies to ensure the argument holds up under scrutiny.
- Clarity: present the reasoning for support of a claim in a clear manner, ensuring (to the extent possible) that the logic is comprehensible and accessible to stakeholders with varying levels of technical expertise.

Poorly-made arguments do not imply a bad ADS and well-made arguments do not necessarily imply a good ADS; rather they speak to the ability of the safety case authors to communicate the safety case's contents. Yet, poor argumentation may be a symptom of broader challenges and/or of the lack of clarity into how the determination of safety is carried out and, as such, should not be easily dismissed. A number of format-dependent elements of a safety case can help ensure the robustness and validity of the argument, as explored in the next section.

### *2.4 Other optional, format-dependent elements*

While claims and evidence are recognized as the two required elements that are at the basis for constructing a safety case and clearly shape the argument definition, there are other possible elements that can be used to strengthen said argument. These additional elements are frequently associated with the formatting choice selected for the safety case, such as Goal Structured Notation (GSN), Claims-Argument- Evidence (CAE), or Structured Assurance Case Metamodels (SACM).[2] Based on a literature review of safety case formats, the approach selected for Waymo's safety case can be seen in Table 1. This tabular format is based on an adaptation of the CAE and Toulmin analysis structure (Bloomfield at al., 1998) (Toulmin, 1979) with additional fields for "counter-arguments" and associated "rejections" (referred to as defeaters in UL (2023)), limitations, and a justification narrative (Favarò et al., 2023).

---

[2]Different safety case formats refer to claims and evidence in different terms, for example GSN refers to claims as goals. See Annex C of ISO/IEC/IEEE 15026-2:2022 for a more detailed comparison.

Table 1 - Format employed, with slight revisions of language, in (Favaró et al., 2023)

| Context | # Claims (including supporting decomposition) | | Evidence | Limitations/ Scope [as needed] | Counter Argument + Rejection [as needed] | Justification Narrative |
|---|---|---|---|---|---|---|
| Details on applicable use-case | Claim ID | Individual assertions in support of the top-level goal of safety determination | Link to internal evidence | *Statement of limitations or out-of-scope elements* | *Note possible counter-argument or defeaters of the claim with rationale for rejection of alternatives* | Detailing of the argument in support of 1) high level parent claims or 2) substantiating claims with evidence and other optional supporting elements |

Entries are provided in natural language for internal usability and survivability of the Safety Case itself. The addition of these format-dependent elements can strengthen the argument for claim satisfaction in multiple ways:

- *Counter-arguments* allow for challenging the premise of a claim and provide proposals for alternative approaches that could falsify the claim, demonstrating a level of due diligence and a method to combat potential confirmation bias arising from the use of structured formats (Leveson, 2020).
    - Each counter-argument requires a *rejection* and rationale for why these alternative approaches were not selected or, in other words, an explanation as to why the defeaters are unsubstantiated[3]. Rejections to counter-arguments allow us to pressure test and strengthen the overall argument, including the ability to provide evidence to support the rejection.
- A statement of *limitations*, as suggested in multiple parts of UL (2023), allows the developer to capture existing challenges and helps prioritize future improvements without leading the structure of the argument itself to potentially "oversell" (because of its consistency and formality) the reality of ADS's performance.

Additional elements may enter the argumentation as part of the "assessment of support" process, that is, the act of reviewing and confirming the claims' support through the provided evidence. Within our approach we make use of "*Justification Narratives*" — a prose explanation detailing the rationale as to why all the provided elements, from required evidence to optional counter-arguments and rejections, and limitations, provide support for the claim. Typically, the subject matter expert that is responsible for a particular claim within the safety case (that is, the person who attest to the accuracy of the claim assertion and selects the evidence that is provided to substantiate the claim) will also be required to craft the Justification Narrative, which makes explicit the basis of the argument that connects the claim to the provided evidence and/or child sub-claims.

---

[3] From a logical perspective, this is equivalent to rejecting "not A" (¬A), given the claim assertion A. Within our approach, counter-arguments may reject parts or all of a claim, so that rejections are not considered as a sufficient support mechanism on their own, and need to be complemented by evidence.

# 3. Safety Case Process Flow: Stages and Functions

The elements of a safety case highlighted in the previous section are introduced at various stages along the process of creating, reviewing, and updating the safety case. Clear understanding of such stages is necessary to ensure appropriate management and governance of "building" the safety case as a "product". These stages are represented below in Figure 5. In this section, we present a brief description of each phase, highlighting roles and responsibilities for the various stakeholders involved. Beyond the definition of clear roles and responsibilities (for example, through a RACI chart), considerations on the level of independence needed across each stage shapes governance of the safety case.

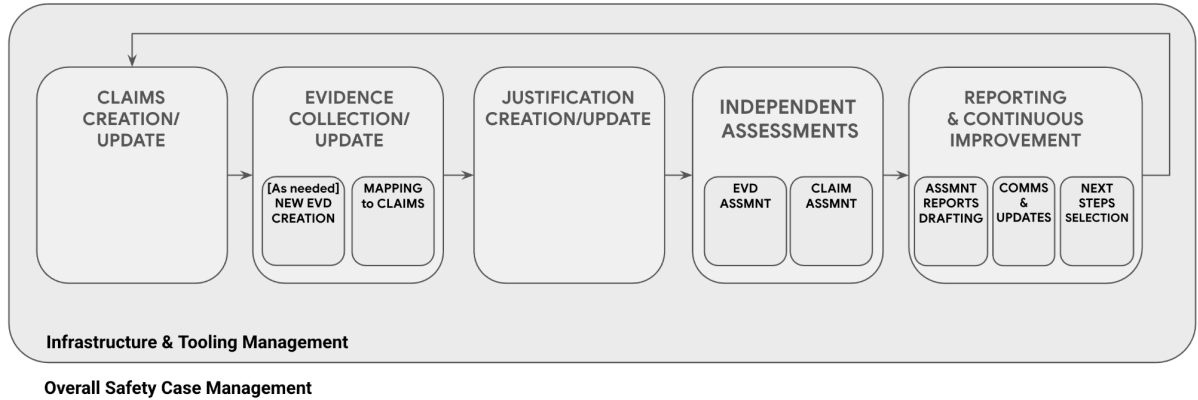

Figure 5. Representation of the safety case process flow.

At a high level, the procedural flow for developing and updating the safety case is defined by the following stages:

- Claims Creation: as presented earlier in the paper, the definition of claims (most often achieved through the methodical decomposition of a top level goal) anchors the entire structure of the safety case. There are multiple approaches that can help drive this decomposition. The creation of claims should be based on an agreed upon template. Said template detailed top-level branches and topical areas of interest, which were identified by abstracting and organizing internal engineering documents and data, as well as industry standards that codify state of the art external expectations. A high level summary of the template was provided above in Figure 2.[4]

- Evidence Collection: Following the creation of claims, collection of artifacts that can substantiate each assertion begins. In Figure 5 we note that, in certain cases, claims may require the ad-hoc creation of evidence that was not originally available. This is considered uncommon though, with the vast majority of claims being supported by both internal and external pre-existing documentation and data that supports the validity and rigor of our approaches.

- Justification Narratives Creation: Once evidence has been appropriately attached and mapped to the relevant claim(s), a prose explanation detailing the rationale as to why the

---

[4] An analysis of sufficiency of a safety case is out of scope for this paper. We note, however, that the chosen structure mimics and trails the internal approval process for each candidate software release for an ADS, so that the claims generation is steeped in the understanding of the internal engineering documentation that form the basis of an ADS Safety Framework (Webb et al., 2020; Favarò et al., 2023, Favarò et al., 2025).

- linked evidence(s) or sub-claim(s) were selected and how they support the claim is also appended to each claim. In other words, the Justification Narrative makes explicit the basis of the argument that connects the claim to the provided evidence or sub-claims. Because of the hierarchical nature of the argument decomposition, claims may be supported by children sub-claims without the need for additional evidence — something that would be noted in the Justification Narrative of the parent claim support. As noted before, claims may be supported by "rejections" to epistemic defeaters — something that would also be noted in the Justification Narrative as a "rejection to possible counter-arguments" that (if applicable) are appended to the claim to strengthen the argument.
- Independent Assessment (scoring): We distinguish between two separate layers of assessment: 1) an assessment and scoring of the claim support through the provided evidence or sub-claims; 2) an assessment and scoring of each individual piece of evidence included within the safety case library (regardless of its mapping to specific claims). Many options exist around the independence of those providing evidence to substantiate each claim and those reviewing the output of such a process. Subject matter experts (SMEs) for each topic covered by a claim are assigned and are responsible for providing evidence, or appointing someone else to, and for creating the associated justification narrative; a team of independent and certified[5] Safety experts would then audit the provided support and carry out the assessment (i.e., assign scoring per the criteria of Tables 1 and 2, covered in more detail below, and provide an assessment summary).
- Reporting and Continual Improvement: Results from the assessment, focusing on all claims not providing adequate support, are then summarized and reported out to company leadership and topic-specific stakeholders through a *safety case assessment report*. Stakeholders are periodically kept informed of relevant updates (e.g., flagging any new assessment report being released) through newsletters and presentations. Together with a summary of the assessment, each report contains suggested actions to be considered toward continual improvement of the safety case as well as of the various related engineering and product workstreams. A cross-functional set of stakeholders meets to review and align on any next steps required in connection with updates for the safety case. Any decision for *updates* (of either overall process or to claims or evidence), would then re-enter the process flow of Figure 5 following the preceding stages highlighted in the prior bullets.

The overall process flow lives within a designated content control environment.[6] This is noted in the background grey box of Figure 5 labeled "Infrastructure and Tooling Management" within which all stages just described are enclosed.

Finally, overarching to the entire process and tooling management is the overall role of oversight of the safety case project. The safety case fulfills an important safety assurance function within the broader SMS. It helps document and pressure-test the integrity of the various engineering and product workstreams that form the basis of approval for a new release of the ADS (Favaró et al., 2025). Other use-cases also exist for a safety case within an SMS, including its direct use toward approval. While possible, that is not the use intended here, where the results and output from this particular assurance activity helps ground continual improvement and maturity of internal safety processes associated with the SMS, and as a key input into overall SMS governance.

---

[5] Certifications included UL-CFSP and UL-CASP:Autonomous.
[6] Within Waymo, JIRA is employed - the same tool used for internal requirements management.

Appropriate governance for the safety case includes having clear documentation and policies that detail each of the stages presented in this section. Within this paper, and the next section in particular, we focus on the fourth step, that is the assessment of the safety case support.

# 4. Assessing a Safety Case: Bottom-up Process

As covered before, the goal of a safety case is to provide a "***[...] compelling, comprehensible, and valid [...]***" argument that a system is safe for a given scope. It is then the job of a reviewer to assess that the argumentation is in fact logical, credible, and well-supported as part of the case credibility assessment. This process starts with the two independent pillars of assessment shown in the fourth box in Figure 5:

   i. an assessment and scoring of the claim support through the provided evidence;
   ii. an assessment and scoring of each individual piece of evidence included within the safety case library (regardless of its mapping to specific claims).

We refer to these two layers as the "assessment of claim support" and the "assessment of evidence status" respectively, and refer to them collectively as the "assessment of case support" as mentioned previously. In this section we outline general guidance to safety case assessors for evaluating both claim support and evidence documentation practices as part of the *safety case assessment phase* that follows the evidence collection step of Figure 5. The intent of this guidance is to ensure high-level consistency in the safety case assessments, as well as align clear expectations between all the stakeholders involved in the creation, evaluation, and usage of a safety case.[7] These two layers of assessment constitute a necessary first step toward the overall determination of credibility of the safety case, but they are not sufficient to establish full credibility of the case on their own. They are part of what we term a "bottom-up" approach to the safety case assessment (see Figure 4). Following their presentation, we describe additional components of a "top-down" assessment within the Limitations section.

## *4.1 Assessing Claim Support*

As noted within the presentation of evidence elements, we distinguish across two types of artifacts: (i) those providing procedural documentation; and (ii) those providing implementation evidence. As such, support for claim is evaluated according to the same two dimensions:

  a. The evaluation of "procedural support" for the claim assesses the extent to which the provided *procedural evidence* articulates a process that comprehensively addresses all aspects of the claim;
  b. The evaluation of "implementation support" for the claim assesses the extent to which the provided *implementation evidence* demonstrates the application of relevant processes, providing results and/or use-case examples that comprehensively address all aspects of the claim.

---

[7] It is understood that some subjectivity is intrinsic in the judgement and expertise of each safety case assessor and, thus, a natural part of their assessment. While this guidance aids consistency in our process, it should not be intended as a prescriptive specification for scoring, nor as suggesting that a conclusive unique assessment is always possible.

The evaluation of a claim support also considers a number of other elements that are specific to a developer's formatting choice for its safety case. The claim support takes into account the statement of rejections to counter-arguments, limitations, and the justification narrative, all of which were introduced earlier in the paper. The justification narrative, in particular, details why a claim is sufficiently supported by evidence or through its sub-claims, without the necessity for attaching evidence artifacts directly to the parent claim.

Based on the body of evidence attached to the claim and its justification narrative, including the (optional) counter-argument and rejection, limitations, and assumptions, the assessor provides two separate scores to evaluate the procedural support and the implementation support for the claim. Each score ranges from 0 (lowest) to 3 (highest), according to the guidance summarized in Table 2 below.

Table 2. Overview of the "claim support" scoring criteria.

| | 0<br>Insufficient Support | 1<br>Initial Support | 2<br>Adequate Support | 3<br>Strong Support |
|---|---|---|---|---|
| **Procedural Support** | Documentation on processes/ procedures has not been provided for any aspects of claim.<br><br>[Procedural evidence may not have been provided, or may be irrelevant to the satisfaction of the claim.]<br><br>No governance or oversight has been provided for any meaningful aspect of the claim. | Documentation on processes/procedures has only been provided for certain portions of the claim and/or is inapplicable to critical aspects of the claim.<br><br>Little or no oversight of processes/procedures has been provided for critical aspects of the claim (for example: reliance on individual or small teams efforts for process design without an approved policy or design document). | Documentation on processes/procedures addresses core aspects of the claim, but may necessitate further clarity or detail.<br><br>Oversight and governance is established, with design aspects and/or procedural decisions under the oversight of key stakeholders at Director level or higher (or clearly defined delegate). | Clear and complete documentation on processes/procedures is sufficient and is directly applicable to all aspects of the claim.<br><br>Oversight and governance is established, with design aspects and/or procedural decisions clearly aligned with broader company programs and goals (e.g., codified in company-level Objectives and Key Results (OKRs) or Product Requirement Documents (PRDs). |
| **Implementation Support[8]** | Evidence of implementing any parts of the processes/ procedures above has not been provided.<br><br>When a process is not specified or required*, evidence of results or metrics relevant for any aspects of the claim is not available.<br><br>[Implementation evidence may not be attached, or may be irrelevant to the satisfaction of the claim.] | Evidence of implementation of processes/procedures is sufficient for some, but not all critical aspects of the claim [it may not have been provided or may be irrelevant].<br><br>Process implementation is ad-hoc, dependent on individual effort, and/or not consistent with documentation provided. When a process is not specified or required*, execution metrics and/or overall results may not support critical aspects of the claim. | Evidence of implementation of processes/procedures addresses core aspects of the claim, with applicable metrics and results that adequately support the claim.<br><br>Process implementation may demonstrate minor inconsistencies or known departures that are not always documented. When a process is not specified or required*, execution metrics and/or overall results still address core aspects of the claim. | Evidence of implementation consistently and appropriately follows processes/procedures documentation, covering all aspects of the claim. Uniform tracking of non-conformities is available and well documented.<br><br>Process implementation demonstrates a deliberate allocation of resources to focus on process optimization & improvement. Applicable metrics are aligned to company goals with tracked progress and trends toward continual improvement. |

---
[8] Depending on the claim, there may be situations in which implementation evidence is not required (e.g., for claims which state the existence of policy or process documentation).

*A safety case aids maturity and defensibility of the approach taken for the determination of safety. While such goals are grounded in the identification and tracking of complete, consistent, and correct documentation, there may be situations in which procedural documentation is not required. For example, claims connected to evidence of clear communication with relevant stakeholders may be sufficiently supported by proof of such communication. The definition of ad-hoc procedural documents (especially after the fact) is not the intent of the safety case, nor is there a desire to create an expectation of procedural documentation for the sake of documentation.*

The development of the claim scoring table is adapted from a variety of maturity assessment models[9]. Disambiguation across the various scoring levels across Table 2 is anchored on the following attributes:

I. Coverage (e.g., call outs of relevance of the provided evidence, from "no" aspects of the claim, to "critical", "core" and, finally, "all" aspects of the claim);
II. Relevance (i.e., a check of the actual applicability of the provided evidence to the given claim); and
III. Governance (e.g., call outs on the level of oversight in the design and implementation of processes, and alignment with stated company objectives and key results).

While not an exact match, these criteria are generally aligned with those called out within (ISO, 2019) in relation to the means for "obtaining and managing evidence".

Table 2 deliberately provides the independent assessor with discretion in how to determine scoring values. The goal of the assessment guidelines is to ensure an appropriate level of consistency across the claims made in the safety case and across different assessors, and not be prescriptive about technical solutions made for each claim. This is consistent with broader safety assurance activities.

The assessment of claim support is based on the body of evidence and/or support of sub-claims, as well as counter-arguments and rejections, limitations, and assumptions, if applicable. The guidance in Table 2 is applied to evaluate the robustness of the justification provided by the claim's Point of Contact (POC).[10] A few additional considerations need to be taken into account for specific types of assessment.

First, when considering multiple pieces of evidence attached to the same claim, the assessor should also evaluate the consistency and coherence of the evidence; questions such as: "Are the findings from different pieces of evidence consistent and supporting each other? Are there any discrepancies, inconsistencies, or contradictions that could weaken the overall argument?" are an integral part of the assessment, even though they are not explicitly stated in Table 2.

Second, when considering the support of a parent claim by its child claims, the assessor should exercise conservativeness when evaluating how the assessment scores for each of the child claims contributes to an overall score for the parent. The assessor should exercise their best judgement to evaluate how much the parent support rests on each child claim, where, in general, there is no precise rule dictating that each sub-claim carries the same "logical weight" toward the support of the parent. This is part of what we term the *qualitative weighting*[11] of different sources for the overall claim support (e.g., evidence at the parent level, contribution from sub-claims, contribution from rejections to counterarguments). Qualitative weighting enables a deeper interpretation of the relative contribution of each claim to the overarching health of the safety case. It avoids, but does not preclude, the use of a strict mathematical averaging of scores, which, while an intuitive practice, can lead to incorrectly weighting of less impactful claims that are not critical (see Table 2) in their contribution to the overall assessment.

---

[9] See the CMMI Institute's Capability Maturity Model Integration (CMMI, 2025), and ISO/IEC 33000 family (ISO, 2015).
[10] Claim POCs are subject matter experts in their respective area.

In general, a review of the justification narrative for the claim is always the starting point for an assessor to construct a holistic picture of the main sources of logical support for the claim. The considerations above influence how the assessor documents their reasoning for each score. In fact, the assignment of scores (one for procedural support and one for implementation support) is complemented by a summary explaining and memorializing which factors drove such selection. The assessor may reference, at their discretion, specific evidence or arguments that support the assignment of one score over another. However, there is not a strict requirement to recount the entirety of the thought process that led to a particular score. The assessor may detail which aspects of the claims were not sufficiently supported by the provided evidence or sub-claims as appropriate, as well as recommend additional evidence to be considered. As needed, group sessions across different assessors can be hosted, to pressure-test and achieve consensus on scoring techniques and establish a robust community of practice and improve inter-rater reliability.

## *4.2 Assessing Evidence Status*

The second component of the safety case assessment is the evaluation of each individual piece of evidence included within the safety case evidence library (regardless of its mapping to specific claims).[12] Evidence is assessed in order to determine how it is managed or controlled for overall maturity (e.g., clear ownership, recency and active/deprecated status, editing/review governance). In this case, scoring ranges from 0 (lowest) to 3 (highest), according to the criteria specified below in Table 3 and following the decision flow visualized in Figure 6. These guidelines apply regardless of the type of evidence (procedural or implementation).

---

[11] Note that qualitative weighting is never intended to counter sound conservativeness practices, nor to create the opportunity to bypass the reporting of low scores for any child claims. Rather, it gives the independent assessor the ability to narrowly point to improvements needed in lower level claims, but still provide an indication of an overall branch acceptability. It also takes into consideration that in the decomposition of claims, owners for parent claims and lower level child claims may be different, and some may provide stronger argumentation and evidence than others.

[12] The scoring of evidence is independent from the claims assessment score. For example, a piece of evidence that exceeds expectations (scored at a 3), could still result in a claim scored, for example, at 0 or 1 should the evidence piece be not entirely relevant and/or not providing sufficient support for core aspects of a claim.

Table 3. Overview of the "evidence assessment" scoring criteria.

| Evidence Assessment | Description |
|---|---|
| 0 - MISSING | No evidence available at this stage, as confirmed by an identified point of contact (POC) for the claim. |
| 1 - DOES NOT MEET EXPECTATIONS | Evidence that is out-of-date and in need of significant revisions (as flagged by the evidence POC), or that is older than 12 months with no time-stamps reconfirming its status as "active". Alternatively, in the presence of a recency check within the past year, evidence that does not showcase clear ownership or whose owner is no longer affiliated with the company/traceable for questions. |
| 2 - MEETS EXPECTATIONS | Evidence that is:<br>● Not older than a reasonable period (e.g., <12 months old since last review);<br>● Time stamped with a review check ensuring the documentation is still active;<br>● Has an owner listed that is still affiliated with the company.<br>Documentation may contain partially out-of-date elements provided they are appropriately flagged. |
| 3 - EXCEEDS EXPECTATIONS | Evidence is up to date (<6 months since review), with current author with active ownership, documented reviews, approvals, and revision history. Evidence is maintained in a controlled environment (e.g., content management system/tooling, etc). |

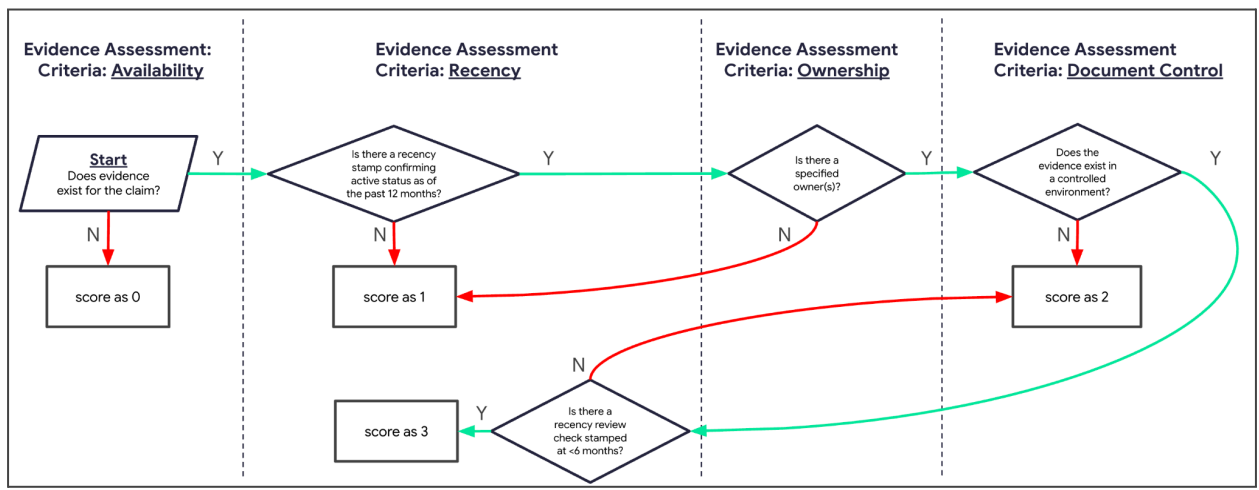

Figure 6. Simplified visual flow for assigning evidence assessment scores.

The evidence assessment also provides an important signal to the broader SMS of the safety case's documentation and records management quality while helping to ensure consistency and quality of the safety case across different teams and divisions that contribute to the overall safety determination. All evidence artifacts should be clearly referenced and linked to the specific claims they support. This traceability ensures that the argument can be easily followed, reviewed, and enables collaboration among stakeholders. It also facilitates the continual updating of evidence throughout the system's life cycle.

### *4.3 Reporting and Continual Improvement*

The output of the assessment phase is a report of results and findings, including quantitative scores associated with both support of claims (part of either individual or aggregated arguments) and documentation status and hygiene of the evidence library. Results and overarching scores can be reported in a number of different formats to aid decision-making. For example, the template structure presented in Figure 2 supports the presentation of results in Figure 7. Each spoke of the plot represents a family/sub-topic of claims according to the argumentation template, and aggregates results from the supporting child claims. Concentric circles along each spoke represent the 0 through 3 scoring established by the guidance of Table 2, to provide a summarizing visual.

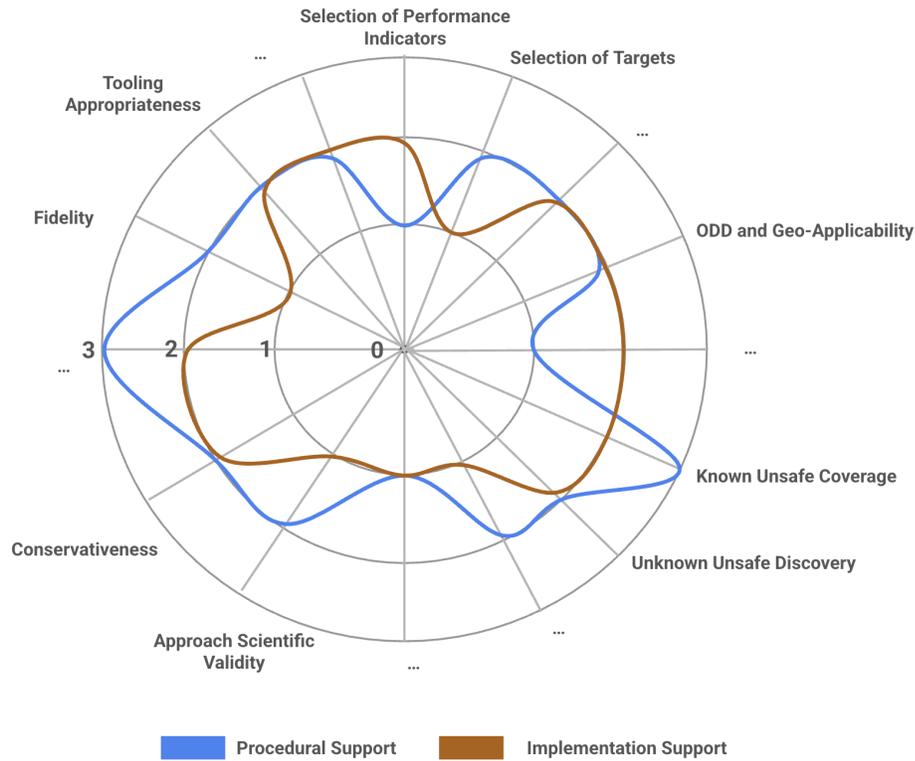

Figure 7. A possible visualization of results, aggregated based on the argumentation template presented in Figures 2 and 3

Reporting of results and findings is primarily used for internal decision-making on continual improvement efforts. A desirable aspect of this reporting is the ability to identify and prioritize trends and themes that may have a higher material relevancy for safety performance.

The escalation of notable, low scores for individual claims or branches of the safety case that require further investment in documentation rigor and/or overall performance, informs the overall credibility and validity of the safety case, but cannot, on its own, support the entire credibility assessment, as further detailed in the limitations of Section 6 below.

## 5. Timing Aspects: When to Write and Assess a Safety Case?

Having covered the safety case elements, process flow, assessment guidelines, and reporting and continual improvement above, the next logical question is pinpointing any guidance around when this process should be initiated and how often updates and the resulting assessment should be implemented. The answer to this question depends on the particular usage that a developer foresees for its safety case. As mentioned previously, the safety case fulfills certain safety assurance functions as part of the broader SMS; to query and investigate the appropriateness of the approach that anchors approval and eventual deployment of a new candidate software release for the ADS (Favarò et al., 2025). The safety case explains and justifies how the approval is undertaken, but it is separate from the approval itself. It follows a parallel workstream that can inform and provide confidence to both internal leadership and external stakeholders about the rigor undertaken in evaluating a new candidate release, but it remains separate from such qualification. Because of this usage, its creation naturally trailed the internal alignment on how the approval process was going to be structured. As noted in Favarò et al. (2023), safety case timing considerations are thus intertwined with company and product maturity aspects. An ADS Safety Framework and its supporting methodologies, like those presented in Webb et al. (2020) afford a level of maturity - not just of the ADS technology, but also of the evaluative approaches used for its qualification - that make the creation of a safety case possible. Maturity and stability of both product and processes make the elucidation and organization of the safety arguments the natural continuation of those internal long-standing safety processes. As both a company technological solution and supporting evaluation approaches mature, a level of stability in overall safety processes is reached that allows to more systematically organize the mapping between safety goals and the evidence we build through the methodologies employed. The SMS in general, and the Safety Case in particular, support safe scaling of a product and steer toward continual improvement and increased maturity, minimizing fragmentation and loss of internal documentation. These concepts are presented in Figure 8.

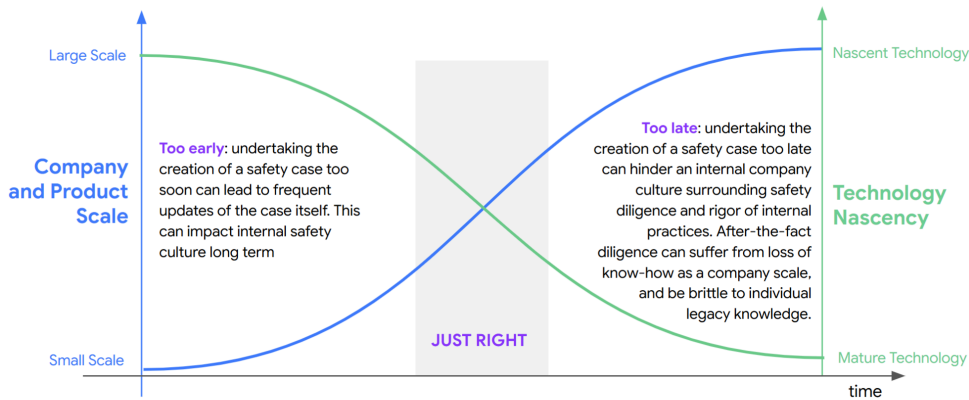

Figure 8. Trade-offs surrounding timing considerations for the creation of a safety case.

Beyond addressing the question of "are internal practices mature enough to support a sufficiently stable safety case" (see the left of Figure 8), one must also consider how frequently the safety case would require updating and re-assessment. This is where appropriate governance can play an important role in shaping how impactful the feedback from each safety case assessment can be to the broader company. Similar to the considerations of Figure 8, there are trade-offs between how frequently a company updates its argumentation (which can result in changes/updated to claims and evidence and the need to re-assess claim support, or - at a minimum - reconfirm the previous assessments) and the ability to provide clear guidance to internal SMEs toward prioritized, meaningful improvements to internal documentation and practices. A cost-benefit analysis might compare the cost and resource pressures of an approach which makes a wholesale update to an approach which prioritizes incremental sustainability of the work product to ensure its validity over time (e.g., impact of changes, required specialist skills, potential updating of tools, or any relevant regulatory changes). Nevertheless, both approaches require investment to ensure the Safety Case remains valid.

A number of options exist for potential triggers for updates to any portion of the argumentation. For example, event based triggers, such as substantial hardware, software, ODD, or use-case changes can be leveraged to inform the need to update the safety case. Other intuitive options, such as time-based triggers for periodic re-evaluation, may not actually reflect the intent and scope of a safety case. Over the years, we have learned that deliberate prioritization and collaboration between safety, engineering, and product teams needs to be established to ensure there isn't divergence of engineering practices and the content reflected in the safety case. Safety argumentation should be updated in the presence of substantial changes in the formulation of an acceptance criterion (see the branches of Figure 2) and/or its evaluation. In other words, any internal change that would invalidate the structure presented in Figure 2 and/or substantially change its content would be cause for an update and re-assessment. Conversely, smaller changes or edits (for example, updating an older piece of evidence with a newer version) are instead considered an integral part of the case management itself, and do not amount to an actual new argumentation or new safety case. We learned that setting too rigid of a structure for re-assessment in the presence of only small changes that do not alter the argumentation logic can in fact backfire and be counterproductive to the shared (i.e., company-wide) effectiveness of the safety case.

## 6. Beyond the Mechanics: Limitations and Out of Scope Assessment Components

As presented in Figure 4, there are a number of top-down assessment attributes that need to be covered in order to achieve an appropriate credibility assessment. Those are not specifically tackled in this publication, but, at a minimum, we can recognize the following two attributes for appropriate arguments, which complement the list presented earlier in section 2.3 (Favaro et al., 2023):

- Completeness: an attribute of coverage qualification, where appropriate arguments ensure there are no significant gaps or omissions in the claims and sub-claims structure. Exhaustiveness of an argument cannot be evaluated in the abstract, so that "completeness" is generally used as the term to ensure that a coverage strategy has been specified and, in fact, fully executed on to determine "completeness with respect to a specification".
- Robustness: a robust argument is one that remains valid even if individual argument elements contain flaws or uncertainties. The overall claims made in the safety case should not be overly sensitive to weakness in individual arguments or pieces of evidence. Further strategies for robust claim decomposition are provided in SCSC (2021) and UL (2023).

These qualifications can be found in the top-right corner of Figure 4, and are connected with and impacted by the particular selection of top-level goal for the case for safety, where the approach selected should account for the ability to produce a sound justification. The investigation of sufficiency is a noted limitation of the present paper, but some of the considerations are included within Favarò et al. (2025; 2023 - see section 2.2), in relation to assumptions/questions that the chosen argumentation approach needs to be able to answer. Said assumptions are connected with the sufficiency of the hazards identified and included in a developer's analysis, and the definition of related acceptance criteria.

Additional qualifications that are out of scope are logical coherence and consistency in the safety argumentation. We do not explore considerations about the need to balance the safety case structure to ensure a common level of abstraction (or detail) is consistently followed across all branches of the safety case. Logical coherence also requires the double checking of each compound argument decomposition, to ensure child claims appropriately and sufficiently support the related parents. Some of those considerations are often connected with a rigorous approach to independent layers of review within the creation and drafting of the safety case claims. Within our approach, independence is embedded in multiple aspects of the stages presented in Figure 5. For instance: claims are drafted by a team that is not directly involved in the development of either the ADS as a product or the evaluation methodologies. Furthermore, our approach includes various rounds of independent reviews, from technical reviews to ensure the accuracy of each assertion, to holistic reviews of the decomposition of claims from dedicated consultants. Assessors of evidence status are not authors, owners or reviewers of said evidence. Similarly, the assessment of each claim support is undertaken by experts that are not involved with the development of the ADS or its evaluative methodologies. Taken together, the degrees of independence built into the process of Figure 5 support the *credibility of implementation* pillar of the CCA (Favarò et al., 2023), and support the top-down credibility assessment discussed before that remains out of scope and a noted limitation of the present paper.

## 7. Conclusions

Safety cases are a key artifact to enable ADS developers to provide a credible argument for a mature safety approach. They also serve an important safety assurance function within the broader SMS by documenting and pressure-testing the integrity of the various engineering and product work streams that form the basis for the approval for a new ADS release. The remaining open question is then how does an ADS developer evaluate its safety case? The presented approach builds on and addresses a gap in the state of the art auditing practices for assessment (ISO, 2019; ISO, 2015; VDA, 2020; UL, 2023) by defining a process for the "bottom up" assessment of evidence and claim support. While not conclusive on its own, it provides a starting point toward a comprehensive "Case Credibility Assessment" (CCA) for ADS safety cases. It does so by providing:

- Key elements of a safety case, exploring the interplay of claims, evidence, and arguments.
- A safety case process flow, including stages for claim creation, evidence collection, independent assessment, and continual improvement.
- A process for assessing both the evidence and claims in a safety case

- Criteria for assessing evidence, emphasizing good document management and maturity,
- Criteria for assessing the support of a claim, in terms of both procedural and implementation support, highlighting the coverage, relevance, and governance of the evidence with respect to the claim it is intending to support.
- Scoring strategies and evaluation guidelines for claim support
- A governance structure to ensure the integrity and effectiveness of the safety case is maintained as ADS technology evolves.
- Timing considerations for creating and assessing a safety case, highlighting the importance of aligning with company and product maturity.

As the development and deployment of ADS continues to accelerate, the role of safety cases will become increasingly vital. This paper provides a starting point towards a practical and transparent assessment of ADS safety cases.